# Chaotic Brillouin Optical Correlation Domain Analysis

JIANZHONG ZHANG,[1,2*] MINGTAO ZHANG,[1,2] MINGJIANG ZHANG,[1,2*] YI LIU,[1,2] CHANGKUN FENG[1,2], YAHUI WANG[1,2], AND YUNCAI WANG[1,2]

[1]Key Laboratory of Advanced Transducers and Intelligent Control System, Ministry of Education and Shanxi Province, Taiyuan University of Technology, Taiyuan 030024, China
[2]Institute of Optoelectronic Engineering, College of Physics and Optoelectronics, Taiyuan University of Technology, Taiyuan 030024, China
*Corresponding author: zhangjianzhong@tyut.edu.cn and zhangmingjiang@tyut.edu.cn

We propose and experimentally demonstrate a chaotic Brillouin optical correlation-domain analysis (BOCDA) system for distributed fiber sensing. The utilization of the chaotic laser with low coherent state ensures high spatial resolution. The experimental results demonstrate a 3.92-cm spatial resolution over a 906-m measurement range. The uncertainty in the measurement of the local Brillouin frequency shift is ± 1.2 MHz. The measurement signal-to-noise ratio is given, which is agreement with the theoretical value. © 2017 Optical Society of America

http://dx.doi.org/10.1364/OL.99.099999

Distributed fiber sensing based on Brillouin scattering has demonstrated to be an excellent technology for long-range distributed strain and temperature measurement by utilizing the ordinary single-mode fiber as sensing element. The distributed sensing technology has found extensive applications in the field of aerospace engineering, civil engineering and smart grids.

Brillouin scattering-based fiber sensing can be classified into two types: time-domain systems, in which the pulse signal is used to perceive the temperature/strain along the fiber, and correlation-domain systems, in which the continuous lightwave with the sinusoidal frequency modulation is utilized as the sensing signal. The time-domain systems including Brillouin optical time-domain reflectometry (BOTDR) [1] and analysis (BOTDA) [2] have obvious advantage in the measurement range, but the best spatial resolution is typically 1 m [3], because of the inherent limitation of the phonon lifetime. To improve the spatial resolution of the time-domain systems, numerous techniques such as differential pulse-width pair BOTDA [4], dark pulse [5], Brillouin echoes [6] and many more have been proposed. However, these improved time-domain systems are comparatively complicated and time-consuming. The correlation-domain systems inclusive of Brillouin optical correlation-domain reflectometry (BOCDR) [7] and analysis (BOCDA) [8] can realize the distributed temperature and strain measurement with the spatial resolution of the millimeter order of magnitude [9], however the measurement range is several meters due to the limitation of Rayleigh scattering [7]. To increase the measurement range of the correlation-domain system, many methods like temporal gating scheme [10], differential measurement scheme [11], and so on have been put forward. But these methods considerably increase the complexity of the correlation-domain systems, although improve their performance to some extent.

To optimize the performance of the correlation-domain systems from the light source viewpoint, some new forms of light with the low coherence are employed as the sensing signals, for instance, the continuous lightwave phase-modulated by a binary pseudorandom bit sequence (PRBS) [12, 13] or the amplified spontaneous emission (ASE) of fiber amplifiers [14]. The chaotic laser source is another way to obtain a reduced level of coherence, which is adjustable by controlling the feedback strength [15]. So the high spatial resolution which is theoretically several centimeter and inverse proportional to the bandwidth of the chaotic light can be achieved. Recently, the chaotic light used as the sensing signals in BOCDR systems has been proposed and demonstrated in our laboratory [16]. However, limited by the spontaneous Brillouin scattering of the sensing fiber, the chaotic BOCDR system has a sensing length of 155m.

In this work, the chaotic BOCDA system is proposed to further extend the sensing length based on the stimulated Brillouin scattering (SBS) of the sensing fiber. The distributed sensing along a 906 m fiber is experimentally verified with a spatial resolution of 3.92 cm, which largely improves the performance of the chaotic BOCDR system.

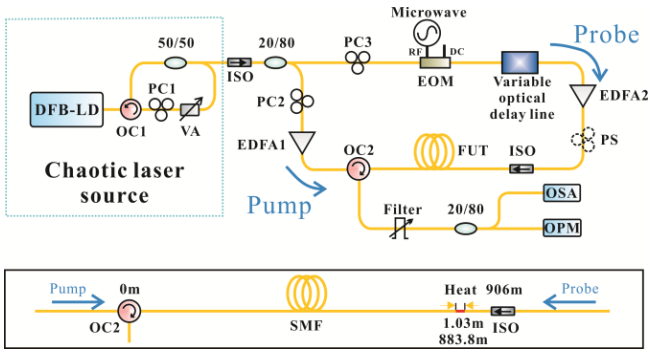

Fig. 1. Schematic diagram of experimental setup for chaotic BOCDA. DFB-LD, distributed-feedback laser diode; OC1, OC2, optical circulator; PC1, PC2, PC3, polarization controller; VA, variable attenuator; ISO, isolator; EDFA1, EDFA2, erbium-doped optical fiber amplifier; EOM, electro-optic modulator; RF, radio frequency; DC, direct current; PS, polarization scrambler; FUT, fiber under test; SMF, single-mode fiber; OSA, optical spectrum analyzer; OPM, optical power meter.

The schematic diagram of the experimental setup for chaotic BOCDA is illustrated in Fig. 1. Chaotic laser source, as plotted in the dashed box, consists of a distributed-feedback laser diode (DFB-LD) and an external feedback cavity formed by the optical circulator (OC1), the polarization controller (PC1), the variable attenuator (VA) and the 3 dB optical coupler (50/50). An appropriate optical feedback is obtained to drive the DFB-LD into chaotic oscillation by adjusting the VA and the PC1.

The output of chaotic laser source passes through the optical isolator (ISO) and then is split into two beams by an optical coupler (20/80). One of the beams (20%), serving as the chaotic pump light, is amplified by an Erbium-doped Fiber Amplifier (EDFA1) after passing through the polarization controller (PC2), and then launched into one end of the fiber under test (FUT) via the optical circulator (OC2). The other beam (80%), as the chaotic probe light, is modulated in suppress-carrier, double-sideband format by the electro-optic amplitude modulator (EOM) which is driven by microwave signal generator. The chaotic probe signal is transmitted through the variable optical delay line, amplified by the EDFA2 and then through the isolator (ISO) launched into the opposite end of the FUT. Besides, a polarization scrambler (PS) is inserted in the probe path for suppressing the polarization dependence of the Brillouin signal. Finally, the chaotic probe light after undergoing the amplification by the chaotic pump along the FUT is filtered by the optical band-pass filter via the optical fiber circulator (OC2). Moreover, 20% of the filtered output is used to monitor the optical spectrum by the optical spectrum analyzer(OSA, Yokogawa AQ6370C), and 80% is connected to a digital optical power meter (OPM, Thorlabs PM100D) with an integrating sphere sensor (Thorlabs S145C) for power acquisition. The structure of the FUT consisting of a 906 m single-mode fiber (G.655), in which a 1.03 m fiber near 883.8 m is placed in a fiber thermostat, is illustrated in the solid line box.

Fig. 2 depicts the auto-correlation trace of the chaotic laser. It can be seen that there is a dominant single correlation peak on the whole curve. Just within this correlation peak, the SBS amplification interaction between the chaotic pump and probe light with the same chaotic state occurs. So, the spatial resolution is theoretically determined by the coherence length of the chaotic light. The coherence length $L_c$ is related to the spectral width $\Delta f$ of the optical spectrum according to the equation of $L_c = c / (\pi n \Delta f)$, where $c = 3\times10^8$ m/s is light speed and $n = 1.5$ the fiber refractive index. The optical spectral width of the chaotic laser is $\Delta f = 2.020$ GHz, as shown in Fig. 3, therefore the theoretical spatial resolution of the chaotic BOCDA system is 3.15 cm. From Fig. 2, we can also see that there are residual side peaks near the main peak. They are resulted by a weak amplitude autocorrelation of the chaotic signal occurring at the delay time of the external cavity [17]. Except the external-cavity-induced side peaks, the autocorrelation sidelobes like the "coding noise" of PRBS exit in the entire trace. These residual off-peak SBS amplification contributes an additional noise mechanism, which can largely limit the signal-to-noise ratio (SNR) of the chaotic BOCDA system. So, averaging over 20 repeating measurements is required to enhance the SNR.

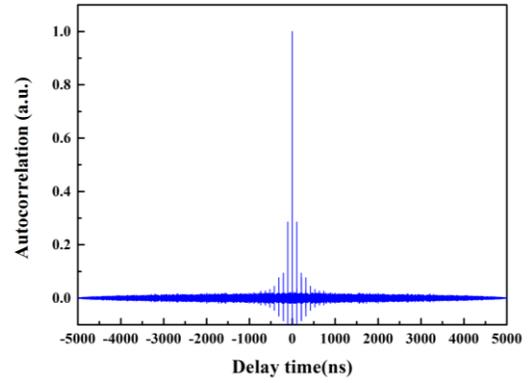

Fig. 2. Autocorrelation trace of the chaotic laser

Fig. 3 shows the optical spectra of different branches in the chaotic BOCDA system. In the pump branch, the optical spectrum of the chaotic light with a center frequency of $v_0 = 1.9303\times10^{14}$ Hz is illustrated as the black curve. In the probe branch, the optical spectrum of the chaotic light with the double-sideband sinusoidal modulation at $v_0 \pm v$ is depicted as the red curve. The modulation frequency $v$ is commonly selected in the vicinity of the Brillouin frequency $v_B$. When the chaotic pump and probe light meet at some location of the FUT, the first low-frequency sideband $v_0$-$v$ is subject to the SBS amplification. The SBS amplification reaches its maximum especially when the modulation frequency $v$ matches with the $v_B$. The optical spectrum of the amplified chaotic probe light is shown as the blue curve. The dark Cyan line in Fig. 3 gives the optical spectrum of the chaotic Stokes light filtered by the tunable optical filter with the bandwidth of 6 GHz.

From Fig. 3, it can be seen that there is 8.74 dB amplification of the chaotic Stokes light in a single acquisition trace. The high amplification mainly results from the accumulation of the residual off-peak SBS amplification instead of the SBS gain within approximate 4 cm correlation peak. This is because that when the chaotic pump and probe light meet in the FUT, the off-peak SBS interaction outside the correlation peak takes place along the entire FUT. Although the generated off-peak acoustic waves are weak and vanish on average, their instantaneous magnitudes are nonzero. So, the off-peak SBS amplification cannot only transfer the part energy of the chaotic pump light to the chaotic Stokes wave, but also the energy of the chaotic Anti-stokes wave to the chaotic Stokes one. As shown in Fig. 3, the gain of the chaotic Stokes wave reaches 6.75 dB due to the energy transfer of the chaotic Anti-stokes light. Actually, the power gain of the chaotic Stokes wave is very weak in the 4-cm correlation peak, on the order of 1.6% for 2 W of chaotic pump power. The temporal gating of the chaotic pump will be employed to suppress the residual off-peak amplification in the future work.

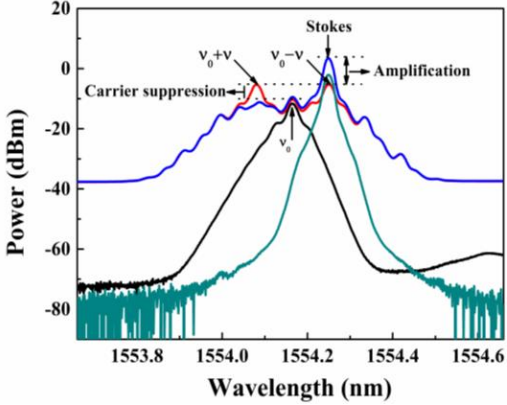

Fig. 3. The optical spectra of different branches in the system. Black line, that of the chaotic pump light; Red line, that of the probe light; Blue line, that of the probe light after SBS interaction; Dark Cyan line, that of the filtered output light.

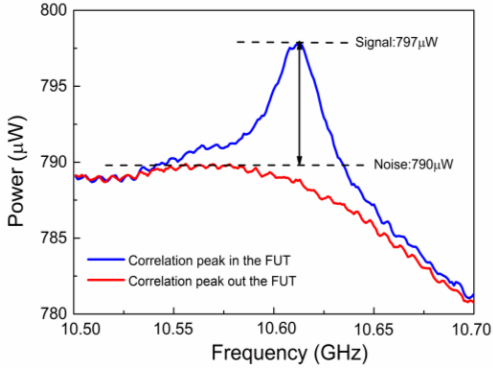

Fig. 4. The gain spectra of the chaotic probe light when the chaotic pump and probe light are injected into the FUT from both ends, respectively. The blue line, that with the correlation peak in the FUT; The red line, that with the correlation peak out the FUT.

The chaotic BOCDA system has similar sensing performance with the ASE-based BOCDA system due to the noise-like characteristic of the chaotic laser. So, learning from a SNR analysis for the ASE-based BOCDA system [14], the SNR of the chaotic BOCDA system can be estimated as:

$$\text{SNR} \approx \frac{1}{2} g_0 \upsilon_g \frac{|A_{p0}|^2}{\Delta f} \sqrt{T \cdot \Delta \nu^{out}} \quad (1)$$

Here $g_0 = 0.2$ [W·m]$^{-1}$ is the SBS gain coefficient of the FUT, $\upsilon_g$ is the group velocity of light in the fiber, $\Delta f = 2.020$ GHz is the bandwidth of the chaotic laser, $\Delta \nu^{out} = 6$ GHz is the bandwidth of the optical filter, and $T = 0.5$ μs is a response time of the integrating sphere power sensor head. $|A_{P0}|^2$ is the mean optical power of the chaotic pump light and on the order of 2 W. Therefore, the expected SNR is approximately 1.09. Fig. 4 illustrates the Brillouin gain spectra (BGS) of the chaotic probe light when the chaotic pump and probe light are injected into the FUT from both ends, respectively. The blue and red lines represent those with the correlation peak in and out the FUT, respectively. The SNR defined as the amplitude ratio of the signal peak to the noise peak [18] is experimentally measured to be 1.01, which is good agreement with the theoretical value.

The BGS in the chaotic BOCDA system is obtained by utilizing a digital optical power meter with an integrating sphere sensor to record the average power of the filtered chaotic Stokes signal versus the modulation frequency $\nu$ [19]. In the experiment, the modulation frequency is swept from 10.5 to 10.7 GHz with the sweep step of 1 MHz.

The length of fibers in the probe and pump branches are matched so that the single correlation peak between the two chaotic waves falls in the FUT. The position of the correlation peak can be scanned over a range of the FUT by a variable optical delay line in the probe branch. It could be composed of the different length cascading fibers, and two programmable optical delay lines, one of which has a large delay range of 0 ~ 20 km with the delay resolution of 30 cm (General Photonics ODG-101) and the other has the more precise delay range of 0 ~ 168 mm with the delay resolution of 0.3 μm (General Photonics MDL-002). Firstly, by adjusting the length of the cascading fibers, the initial position of the correlation peak is located at the end connecting the OC2 of the FUT. Then, the scanning of the correlation peak is accomplished by cooperatively adjusting these two programmable optical delay lines.

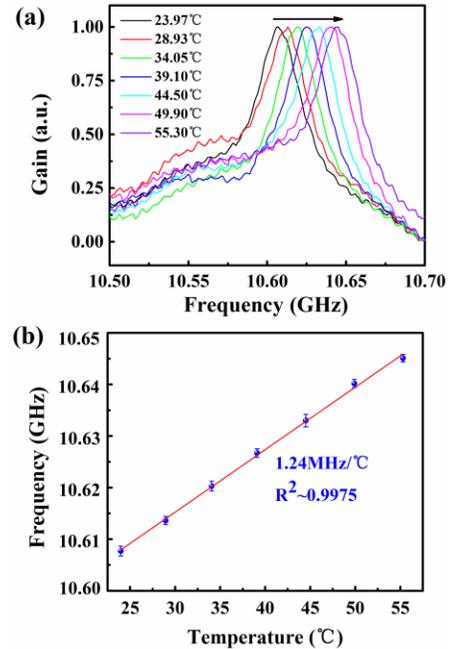

Fig. 5. The relationship of the chaotic BGS with temperature. (a) Temperature-dependence of BGS in the FUT; (b) that of the BFS.

The relationship of the chaotic BGS with temperature is illustrated in Fig. 5. Fig. 5 (a) shows the temperature-dependence of BGS in the FUT. First of all, the single correlation peak of the chaotic laser is located in the fiber thermostat by adjusting the variable optical delay line in the probe path. Then the temperature of the fiber thermostat is changed from 23.97 to 55.30 °C in succession. A phenomenon that the center frequency of the BGS moves from 10.608 to 10.646 GHz can be obviously observed, where the linewidth of the BGS is stabilized at 44.9 MHz. From these BGS spectra, we can plot the temperature-dependence of Brillouin frequency shift (BFS) as shown in Fig.5 (b). According to the fitting curve, the sensitive temperature coefficient of this system is 1.24 MHz/°C. Fig.5 (b) further depicts the uncertainty of the measured BFS at each temperature, showing that the maximum standard deviation is ±1.2 MHz.

Fig. 6 (a) shows a three-dimensional plot of the BGS measured along the FUT. The 1.03-m long heated fiber near 883.8 m can be clearly distinguished. In our experiment, the optical fiber thermostat is set to 55 °C close to the maximum temperature (60 °C) of this instrument and the room temperature maintains constantly at 25 °C. Fig. 6 (b) plots the measured distribution of the BFS along the FUT. The inset of Fig. 6 (b) shows the enlargement of the BFS distribution along the 1.03 m-long heated fiber. The change of the BFS is about 37 MHz, which matches with a 30 °C temperature variation. The spatial resolution of the chaotic BOCDA system reaches 3.92 cm, which can be calculated by the average value of 10%~90% of the rise and fall time equivalent length in meter for the heated fiber section[4]. The experimentally obtained spatial resolution is nearly consistent with the above theoretical one. Actually, the spatial resolution of the chaotic BOCDA system can be further improved to a millimeter or sub-millimeter level by increasing the linewidth of the chaotic laser.

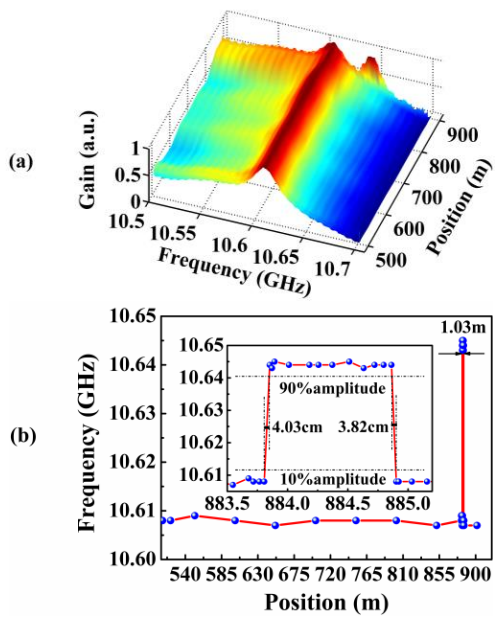

Fig. 6 (a) Distribution of the BGS along the FUT; (b) Distribution of the BFS along the FUT.

In the proof-of-concept experiment of the chaotic BOCDA system, the distribution location of the temperature is achieved by a variable optical delay line, as mentioned before. However, this variable optical delay line is complex in structure and inconvenient in utilization. To overcome this difficulty, for one thing, we are consulting some high-tech companies like General Photonics, whether automatic delay lines with larger scanning ranges and the more precise delay resolution are readily available. For another, we are attempting to adopt chaotic correlation optical time domain reflectometry technology to locate the temperature position along the FUT, which has been applied in the measurement of fiber attenuation [20, 21].

In summary, a novel BOCDA based on the chaotic light for the distributed fiber sensing is proposed and experimentally demonstrated. The BGS of the temperature-dependence is measured by utilizing a digital optical power meter to record the average power of the filtered chaotic Stokes signal versus the modulation frequency. The location of the temperature is scanned along the FUT by a variable optical delay line. Our system can successfully achieve the distributed temperature measurement with 3.92-cm spatial resolution and 906-m measurement range, which can be applied in the aerospace and aircraft industry for structural health monitoring.

**Funding.** National Natural Science Foundation of China (NSFC) (61377089, 61527819); Shanxi Province Natural Science Foundation under Grant (2015011049); Research Project Supported by Shanxi Scholarship Council of China under Grant (2016-036)

**Acknowledgment**. We extend special thanks to Prof. Xiaoyi Bao from University of Ottawa, Canada for valuable discussions.